# THE INFLUENCE OF CRACK-IMBALANCE ORIENTATION AND ORBITAL EVOLUTION FOR AN EXTENDED CRACKED JEFFCOTT ROTOR

J. Gómez-Mancilla [1], J-J. Sinou [2*], V.. R. Nosov [1], F. Thouverez [2], A. Zambrano R. [1]

[1]Lab. Vibraciones & Rotodinamica ESIME, Instituto Politécnico Nacional, México D.F., MEXICO.

[2]Laboratoire de Tribologie et Dynamique des Systèmes UMR CNRS 5513,
Ecole Centrale de Lyon, 36 avenue Guy de Collongue, 69134 Ecully, France.

**Abstract**
   Vibration peaks occurring at rational fractions of the fundamental rotating critical speed, here named Local Resonances, facilitate cracked shaft detection during machine shut-down. A modified Jeffcott-rotor on journal bearings accounting for gravity effects and oscillating around nontrivial equilibrium points is employed. Modal parameter selection allows this linear model to represent first mode characteristics of real machines. Orbit evolution and vibration patterns are analyzed yielding useful results. For crack detection results indicate that, instead of 1x and 2x components, analysis of the remaining local resonances should have priority, this is due to crack-residual imbalance interaction and to 2x multiple induced origins. Therefore local resonances and orbital evolution around ½, $^1/_3$ and $^1/_4$ of the critical speed are emphasized for various crack-imbalance orientations.

Keywords:
Dynamic systems, rotor dynamics, crack detection, local resonance, extended Jeffcott rotor, orbital evolution.

**Résumé**
Les pics de vibration apparaissant au passage des fractions de la vitesse de critique de rotation des systèmes tournants, appelées résonances locales, facilitent la détection de fissures sur les machines. Dans cette étude, un modèle de rotor Jeffcott modifié avec une fissure tournante, comportant des coussinets et prenant en compte les effets de pesanteur et de balourd est présenté. Le choix modal des paramètres permet de représenter les caractéristiques liées au premier mode des machines tournantes usuelles. Les évolutions des vibrations et des orbites du système comportant une fissure sont analysées et permettent d'obtenir des résultats utiles pour la détection des fissures sur les machines tournantes. Ainsi, ces résultats indiquent que, en plus des composants 1x et 2x, l'analyse des autres résonances locales restantes doivent être regardées avec attention du fait de l'interaction possible entre les différentes orientations de la fissure et du balourd, et des origines multiples pouvant engendrer la présence des résonances 2x. Par conséquent les résonances et l'évolution des orbites obtenus autour de ½ , $^1/_3$ et $^1/_4$ de la vitesse critique sont étudiées pour différentes variations d'angle entre le balourd et l'orientation de la fissure.

Mots-clés:
Dynamique des vibrations, dynamique des rotors, détection des fissures, résonance locale, modèle de Jeffcott étendu, évolution des orbites.

## 1. INTRODUCTION

Personal safety, operating costs and increasing overhaul-time intervals motivate research in cracked rotor detection and make structural assessment by monitoring vibration much convenient. Crack detection is most feasible during frequency sweeps such as machine shut-down, when several local vibration resonances occurring at rational fractions of the fundamental critical speed can indicate structural problems. There has been extensive research on the vibration behaviour of crack rotor and the use of response characteristics to detect cracks [1-9].

In his literature review Gash [3] provided a useful survey on the state of the art in this field. Friswell team has work in several interesting topics among which the simplified models for the crack [4] are analyzed. Penny and Friswell conclusion has relevance since the simple harmonic crack breathing function as proposed by Mayes [2] allows this model to be linear and yet yield similar system behavior as an equivalent and more involved nonlinear model. Baschmichdt and Pennacchi [8] have carefully studied and developed crack modeling, mostly when affected by thermal effects as it occurs in vertical cooling pumps. For the case of a vertical cracked rotor having significant gyroscopic effects Yamamoto and Ishida [5] proposed an interesting nonlinear model also capable to analyze the crack-imbalance orientation effect in the absence of gravity terms. Although for quite different rotor configurations both works, [5] and the present one reach similar results concerning the existence of Local Resonances, the masking influence of mass imbalance, and the complex system vibration responses.

In this study, an extended Jeffcott Rotor on lubricated journal bearings having masses and imbalances at disc and bearings developed by Gómez-Mancilla [6-7], is used to characterize orbital evolution and vibration patterns at the local resonances, including the critical speed. The model here used includes gravity and yield periodic responses even in the absence of imbalance excitations. In rotors with small gyroscopic effects where the coupling to other shafts is flexible, and a crack is located near or at the shaft mid-span, the present simplified model is sufficiently for good qualitative description of actual machine vibrating at its first mode. While using Jeffcott-rotor models the key relays on proper selecting the modal parameter values, which allows adequate representation for real machines.

The relative orientation angle between residual imbalance and crack and its interaction drastically affect the system response making crack detection very difficult. Existence of Local Resonances at several rational sub-super and combination harmonics permit good vibration characterization reducing the influence of interaction between unknown vectors; i.e., residual imbalance and cracked shaft magnitudes and orientations. Certain controversy exists while establishing which component is more relevant to damage detection, synchronous 1x or 2x vibration. For practical purposes, crack presence significantly affects the synchronous 1x component and renders confusing prognosis due to its large dependence on the magnitude and on the relative phase angle existing between the unknown residual imbalance and crack vectors. Recall that in real situations crack-imbalance magnitudes and relative angle are not known a priori. Then the authors recommend to focus on the all combinations of 1,2,3 super-sub-harmonic components since all of them are capable of inducing what we call Local Resonances. For the previous reasons evolution and analysis of orbital shapes due to speed sweep with special emphasis on Local Resonances of synchronous, twice and three times harmonic are presented.

## 2. EXTENDED JEFFCOTT ROTOR

In ref. [6] the used mathematical model is introduced with detailed derivation of the equations; and its physical configuration is illustrated in Figure 1. The modified Jeffcott-rotor on journal bearings here used includes gravity, crack and unbalance; then, for self-contained purposes the mathematical equations expressed in dimensionless form and its main characteristics are briefly presented. A total of four degrees of freedom, two at the disc $X_d$, $Y_d$, plus two at both bearing locations $X_b$, $Y_b$ result. Bearing differential displacements, the eight rotordynamic coefficients and the corresponding bearing equilibrium locus $\varepsilon_{xo}$, $\varepsilon_{yo}$, are used to express the bearing reaction forces yielding oscillations around nontrivial equilibrium points.

Applying Newton Second Law at the disc location and normalizing each term in accordance to the nomenclature and using the coordinate system as figure 1(b) we obtain,

$$\Omega^2 \ddot{X}_d + 2D_e \Omega \dot{X}_d + (X_d - X_b) - \frac{g(\Phi)}{2} \cdot \left[ (\Delta K_1 + \Delta K_2 \cos 2\Phi)(X_d - X_b) + \Delta K_2 \sin 2\Phi (Y_d - Y_b) \right] = W_g + \Omega^2 U_d \cos(\tau + \varphi_d) \quad (1)$$

$$\Omega^2 \ddot{Y}_d + 2D_e \Omega \dot{Y}_d + (Y_d - Y_b) - \frac{g(\Phi)}{2} \cdot \left[ \Delta K_2 \sin 2\Phi (X_d - X_b) + (\Delta K_1 - \Delta K_2 \cos 2\Phi)(Y_d - Y_b) \right] = \Omega^2 U_d \sin(\tau + \varphi_d) \quad (2)$$

In a similar fashion, force balance and normalization at the bearing locations yields,

$$\Omega^2 \alpha \ddot{X}_b + \tfrac{1}{2}(X_b - X_d) - \frac{g(\Phi)}{4} \left[ (\Delta K_1 + \Delta K_2 \cos 2\Phi)(X_b - X_d) + \Delta K_2 \sin 2\Phi (Y_b - Y_d) \right] + \left( K_{xx} \Delta X_b + K_{xy} \Delta Y_b + \Omega C_{xx} \Delta \dot{X}_b + \Omega C_{xy} \Delta \dot{Y}_b \right) = \Omega^2 \alpha U_b \cos(\tau + \varphi_b) - W_g/2 \quad (3)$$

$$\Omega^2 \alpha \ddot{Y}_b + \tfrac{1}{2}(Y_b - Y_d) - \frac{g(\Phi)}{4} \left[ \Delta K_2 \sin 2\Phi (X_b - X_d) + (\Delta K_1 - \Delta K_2 \cos 2\Phi)(Y_b - Y_d) \right] + \left( K_{yy} \Delta Y_b + K_{yx} \Delta X_b + \Omega C_{yy} \Delta \dot{Y}_b + \Omega C_{yx} \Delta \dot{X}_b \right) = \Omega^2 \alpha U_b \sin(\tau + \varphi_b) \quad (4)$$

where relevant parameter values are $\Delta X_b = X_b - \varepsilon_{xo}$; $\Delta Y_b = Y_b - \varepsilon_{yo}$; $\Delta X_d = X_d - (\varepsilon_{xo} + W_g)$; $\Delta Y_d = Y_d - \varepsilon_{yo}$; $\Delta K_1 = \Delta K_\xi + \Delta K_\eta$; $\Delta K_2 = \Delta K_\xi - \Delta K_\eta$; $\Phi = \omega t + \varphi_d + \beta = \tau + \varphi_d + \beta$; $W_g = \delta_s/C_r$.

The existence of mass at the disc and at the bearings allows performing several types of useful analyses such as, a number of imbalance combinations varying relative magnitudes and angular phases at the disc-bearing and with respect to the crack orientation; also by varying α different shaft lumped mass distributions can be accounted for. A simple crack breathing phenomenon such as discussed by Gash [3], typical in weight dominated systems, is assumed:

$$g(\Phi) = \left(\frac{2}{\pi}\right)\left(\frac{\pi}{4} + \cos\Phi - \frac{1}{3}\cos 3\Phi + \frac{1}{5}\cos 5\Phi - \frac{1}{7}\cos 7\Phi + \ldots\right) \quad (5)$$

## 3. NUMERICAL RESULTS

A single mid-span crack with medium crack size depth ($\Delta K_\xi = 0.10$ and $\Delta K_\eta = 0.017$) corresponding to 40% of the shaft diameter is used with $U_d = 0.08$, corresponding to similar crack-imbalance interaction. Crack breathing model displaying a squared period, as indicated in Equation (5) is employed; yet as previously mentioned by Friswell [4], results are lightly depended on the breathing model. The important relative angular orientation existing between the disc imbalance vector and the crack ξ-axis, see Figure 1, is varied and computed for various orthogonal directions. Rest of the used simulation parameter values are as follows: relative mass at the bearings, α = 0.50, corresponding to approximate uniform shaft mass distribution; weight parameters, $W_g$ = 2.0, is a weight dominant sag; journal bearing to operating critical speed numbers ratios, So = 0.90 corresponding to relatively flexible shaft-bearing support stiffness ratio. The shaft supported on relatively short journal bearings having L/D = 0.5 and corresponding bearing dynamic coefficients are employed.

In Figure 2, horizontal and vertical vibrating amplitudes at disc and bearings plotted as functions of the operating frequency $\Omega$ are illustrated for various crack-imbalance vectors orientations. Local resonance peaks both, in vertical and horizontal directions show small yet significant differences in magnitude, and peak occurs at slightly different speeds; also clear differences between orthogonal directions and between disk and bearing locations can be observed. The latter differences can be attributed to weight effect and bearing asymmetric characteristics as well as damping existing at the bearings. Moreover, a significant variation both on magnitude and phase angle of the response occurs due to the relative phase angle between the crack and the imbalance vectors thereby impacting on the orbit shapes and its associated evolution.

As illustrated by Figures 2, crack and residual imbalance (both been unknown vectors) can mask the crack's presence and make traditional detection techniques difficult. Unfortunately such interaction mostly affects both, operating speed and twice synchronous frequencies; i.e., 1x and 2x respectively. Since the 2x component is also attributed to misalignment and to other symptoms, the mere presence of 2x does not unequivocally means a crack. Moreover, depending on the relative angle between imbalance and crack vectors $\phi_d$, mostly 1x component can increase or even decrease in magnitude, thereby considerably deform orbit shapes at disc and bearings.

Moreover, disc and bearing orbit responses at some Local Resonances for four crack-imbalance relative orientation angles ($\phi_d = 0$, $\pi/2$, $3\pi/2$, $2\pi$) are presented in Figures 3. In this fashion, two important features of the cracked system can be examined, its orbital shapes affected by the $\phi_d$ angle, as well as the vibration evolution with the operational frequency sweep for the range $0.3 < \Omega < 0.6$, as illustrated in Figures 4 and 5. Then, the classical orbital internal loop at speed $\Omega \approx 0.5$ (mostly due to 2x component) is practically independent of crack-imbalance angular orientation, as illustrated in in Figures 3(b) and 3(d). On the other hand this model also produces outside-inside loops phenomena and orbital angle evolution of the internal loop around 1/3 and ½ of the first resonance, as observed experimentally by by Adewusi and Al-Bedoor [9] (Figures 4 and 5). As expected, the synchronous response component is the most influenced by this angular variation.

Perhaps with the exception of the local resonance around $\Omega \approx$ ½, every other local peak occurring within the range $0.45 < \Omega < 1.20$ displays sufficient dissimilar vibration pattern and orbit; see overlapped orbits in Figs. 3. However all local resonances outside this frequency range, plus the one at $\Omega \approx$ ½, can be properly post-processed and analyzed to reveal the crack presence.

Therefore crack detection at low frequency (sub-critical speed range) is possible, since for typical system parameter values resonance amplitudes at fractions equal/lower than one half the normalized critical speed (i.e., 45%), are in general significant. That is, for medium/larger cracks vibration magnitudes and orbit shapes at lower resonances are generally large enough to be measured, processed and analyzed. In this manner, mid-span crack detection hampering by imbalance-crack interaction becomes much less influential.

## 4. CONCLUSION

An extended cracked Jeffcott-rotor, which applies well to simple machines flexibly coupled and supported on journal bearings where a crack at or near its shaft mid-span exists and having small gyroscopic effects, is used. The model has advantages of linear systems, yet the nontrivial equilibrium approach yields multi-frequency response which allows characterizing cracked shafts.

During run up/down several Local Resonance peaks at fractions of the operating normalized critical speed occur in cracked shaft machines. Vertical and horizontal responses of the disc and bearings orbital evolution around nontrivial equilibrium and Bode plots generated by frequency sweep and by orthogonally varying the imbalance orientation for a rotor configuration having similar crack-imbalance influences are analyzed. It results that orbital evolution around ½ and $^1/_3$ of the first resonance can be used to detect rotor cracks, even if the crack-imbalance orientation is unknown.

**Nomenclature**
$C_r$: bearing radial clearance
$k_{i,j}$ $(i,j = x,y)$: bearing stiffness coefficients; $c_{i,j}$ $(i,j=x,y)$ : bearing damping coefficients
$eu = e0$ : imbalance mass eccentricity; $e_{x0} = e_0 \cos\varphi$, $e_{y0} = e_0 \sin\varphi$
$ks$ : integral un-cracked shaft stiffness
$\Delta k_\xi$, $\Delta k_\eta$ : crack stiffness change, directions $\xi$, $\eta$
$m_d$, $m_b$: concentrated mass, disc and bearing, responses
$W$ : force load on bearing.
$\delta s = mg/ks$ : static shaft deflection
$\omega$ = rotor operating speed [rad/s]
$\omega_c$ = rigid support critical speed
$\xi$, $\eta$ = rotating coordinates, $\xi$ crack orientating axis
$\alpha$ : bearing and disc mass ratio
$U_d$ : disc mass imbalance ratio magnitude *(=eu/Cp)*
$K_{ox}$, $K_{oy}$ : compensating equil. bearing stiffness
$\Delta K_\xi$ : stiffness change along dir. $\xi$ $(= \Delta k_\xi /ks)$
$\Delta K_\eta$ : stiffness change, orthog. dir. $\eta$, $(=\Delta k_\eta /ks )$
$D_e$ : external damping $(= cd / 2 m_d \omega_c)$
$So$ : fixed Sommerfeld number, $= S/\Omega = DL\omega_c\mu / 2\pi W(R/ C_r)^2$ with $S$ bearing Sommerfeld number
$W_g$ : gravity sagging parameter $(=\delta s /Cp)$ with $\delta s = mg/ks$: static shaft deflection, sagging parameter
$W_{brg}$ : $2\alpha W_g$ force load at brg. locations
$\varphi$: angle of imbalance vector *w.r.t.* x-reference
$\beta$ : relative angle between crack and imbalance
$\Phi = \omega t + \beta = \tau + \beta$ : instantaneous rotating angle

**Acknowledgments**
The first author is grateful to *Ecole Centrale de Lyon*, France, for financial support during his sabbatical year. Work partially sponsored by *Consejo Nacional de Ciencia y Tecnologia, CONACyT*, Project 38711-U. Thanks to *S.N.I.* and *EDI* scholarships granted by *CONACyT*, and *Instituto Politécnico Nacional, IPN*, respectively. Graduate student Zambrano upgraded program for this study.

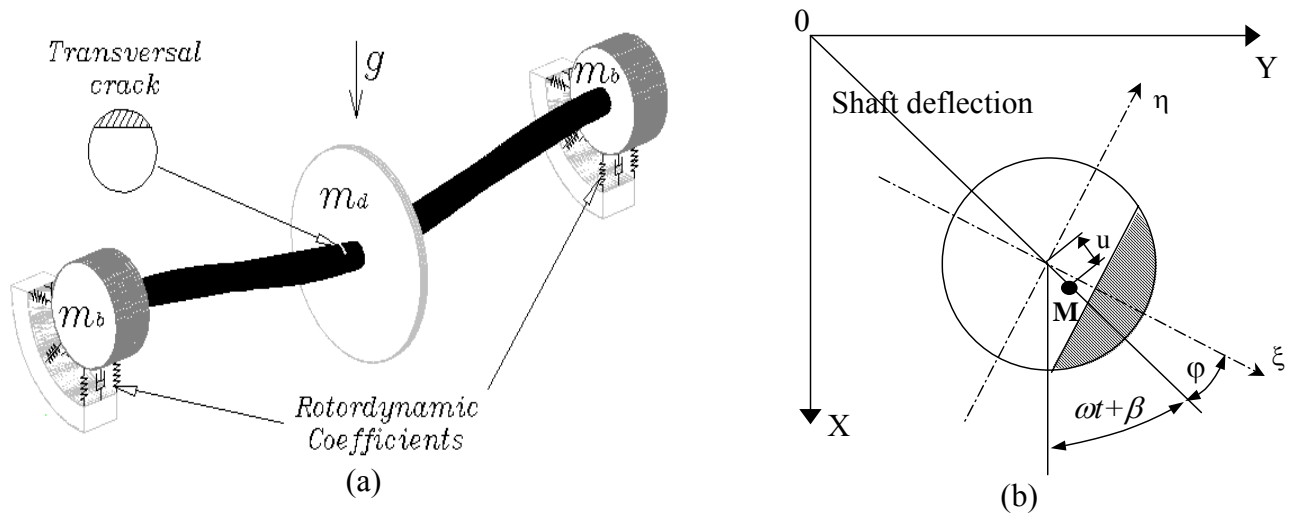

Figure 1: Modified Jeffcott-rotor showing crack location; gravity load; imbalance and masses at both, disc $m_d$, and bearings ends $m_b$; bearing rotor-dynamic coefficients.
(a) Cracked rotor-bearing system; (b) coordinate system crack and disc imbalance

Figure 1: Jeffcott rotor étendu montrant la position de la fissure, le balourd, les masses au niveau du disque et des coussinets et la force de gravité
(a) Système rotor-coussinet-fissures (b) Coordonnées du systèmes fissuré et du balourd associé

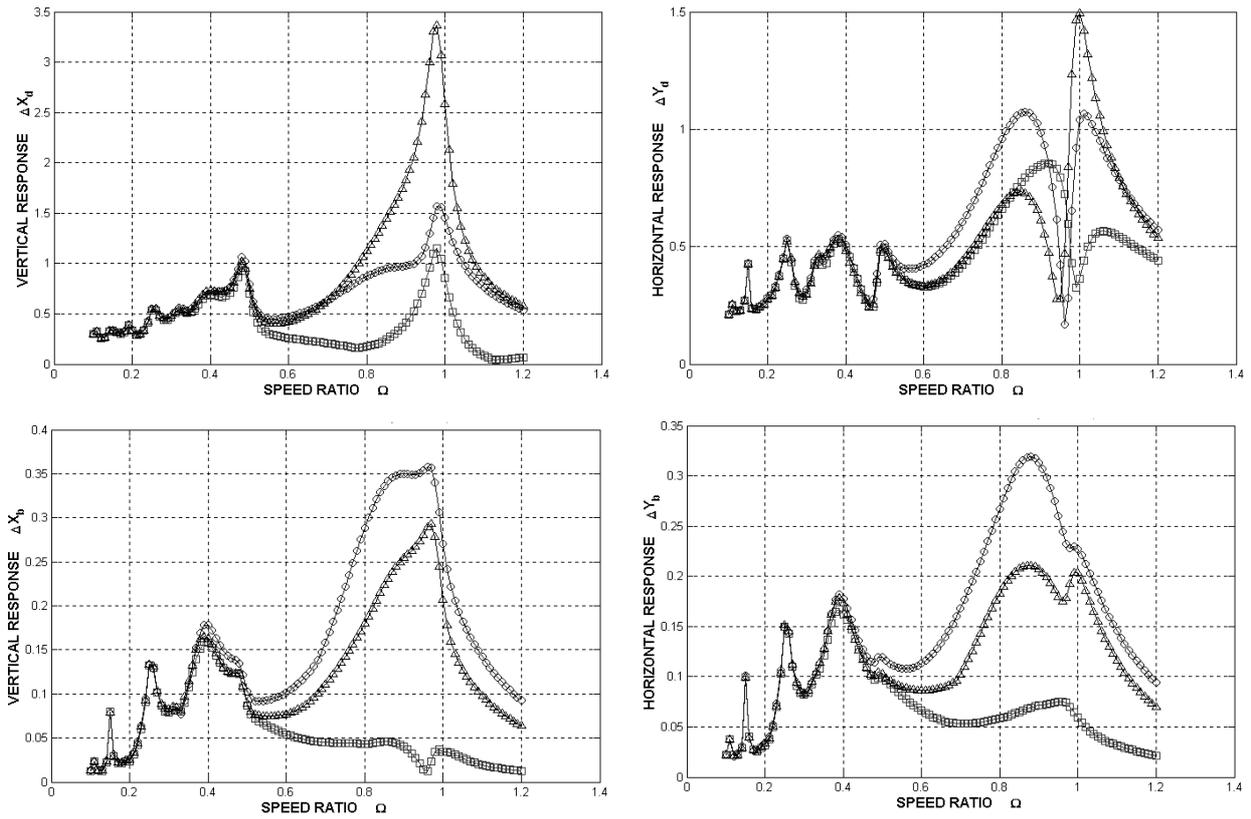

Figure 2: Vertical and horizontal responses of the disc and of the bearing for various crack-imbalance orientation
($\Phi_d = 0°$: circle, $\Phi_d = 90°$: square, $\Phi_d = 270°$ : triangle)

Figure 2: Responses verticale et horizontale du disque et des coussinets pour différentes valeurs d'orientation entre la fissure et le balourd
($\Phi_d = 0°$: cercle, $\Phi_d = 90°$: carré, $\Phi_d = 270°$ : triangle)

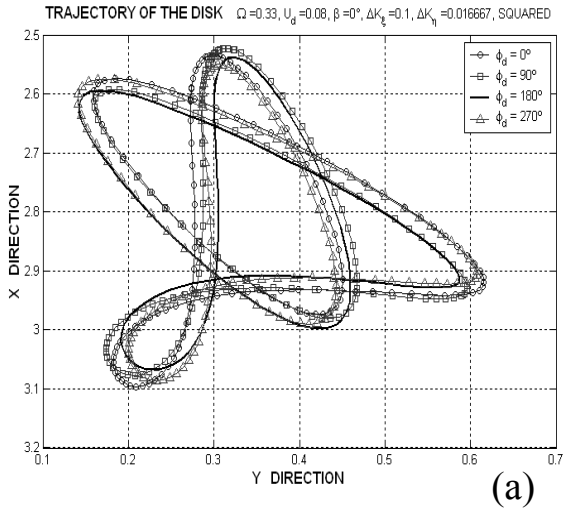 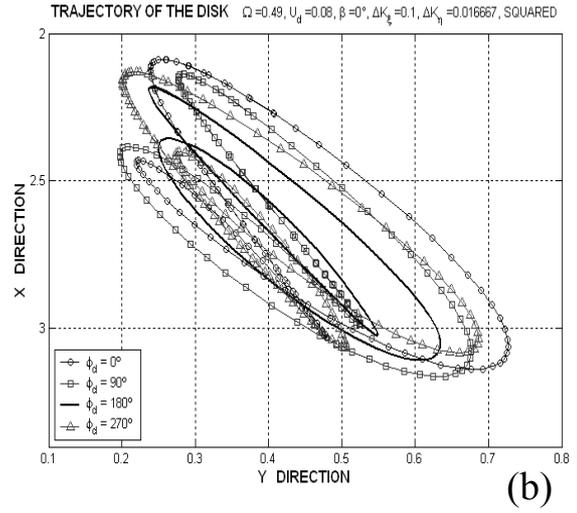
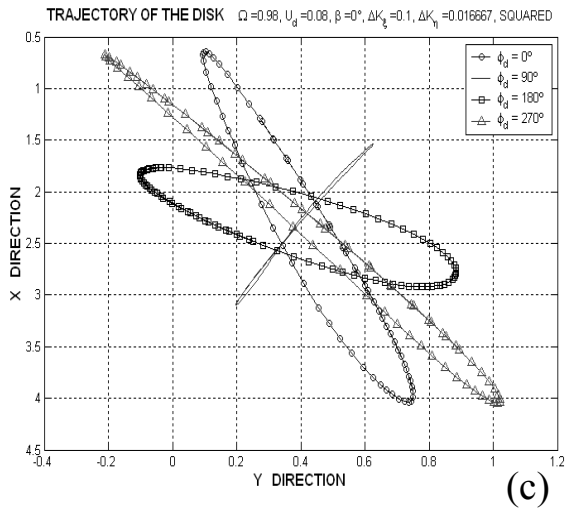 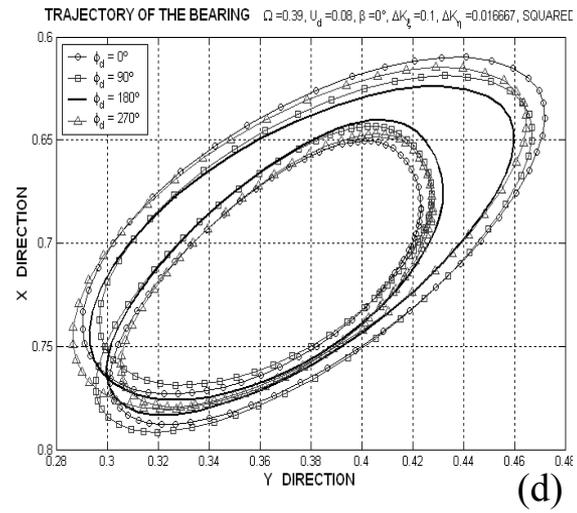

Figure 3: orbits at some local resonances for various imbalance-crack orientations $\phi_d = 0, \pi/2, 3\pi/2, 2\pi$.
(a) trajectory of the disk for $\Omega=0.33$ (b) trajectory of the disk for $\Omega=0.49$
(c) trajectory of the disk for $\Omega=0.98$ (d) trajectory of the bearing for $\Omega=0.39$

Figure 3: orbites à différentes résonances locales pour des différentes orientations de fissure et balourd $\phi_d = 0, \pi/2, 3\pi/2, 2\pi$.
(a) trajectoire au niveau du disque pour $\Omega=0.33$ (b) trajectoire au niveau du disque pour $\Omega=0.49$
(c) trajectoire au niveau du disque pour $\Omega=0.98$ (d) trajectoire au niveau des coussinets pour $\Omega=0.39$

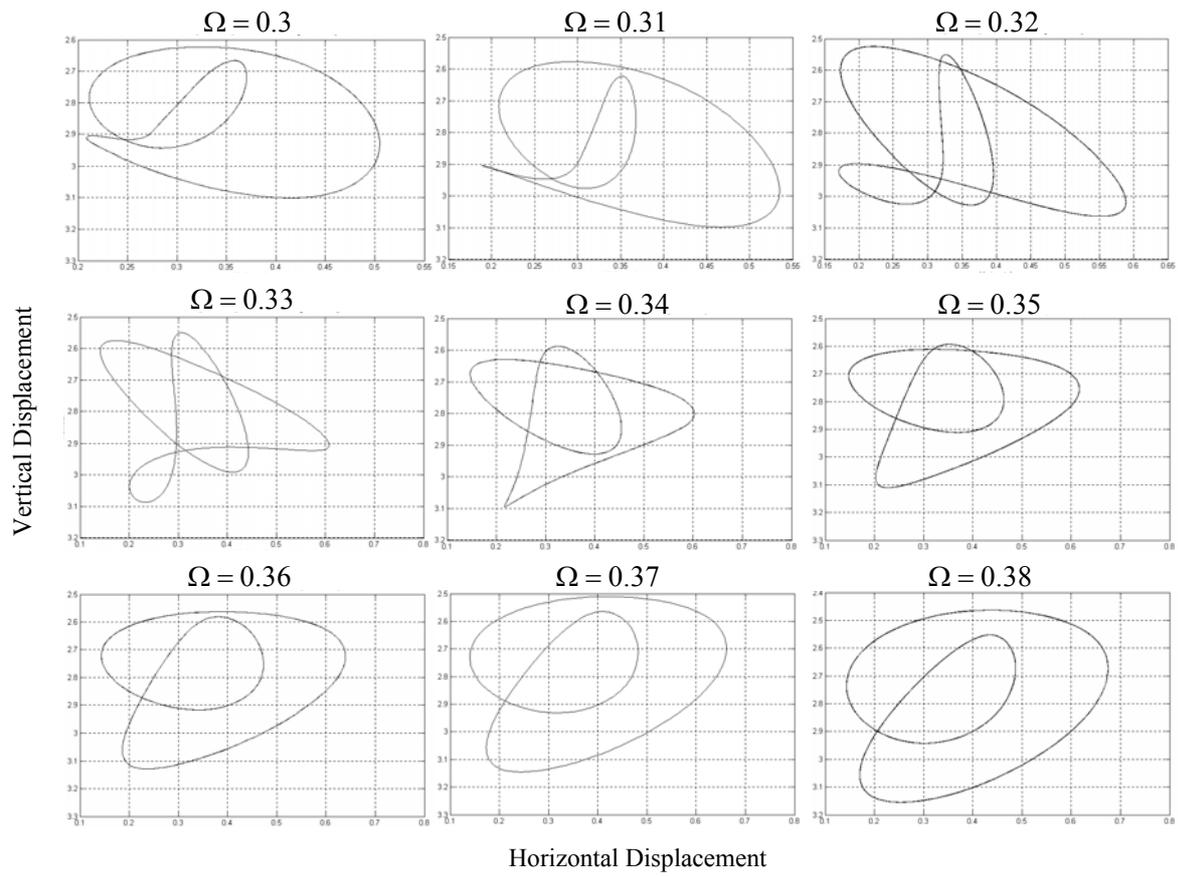

Figure 4: Orbital evolutions of the disc around one third of the resonance
(Crack-imbalance orientation $\Phi_d$ =270°)

Figure 4: Evolutions des orbites au niveau du disque autour du tiers de la résonance
(Orientation fissure-balourd $\Phi_d$ =270°)

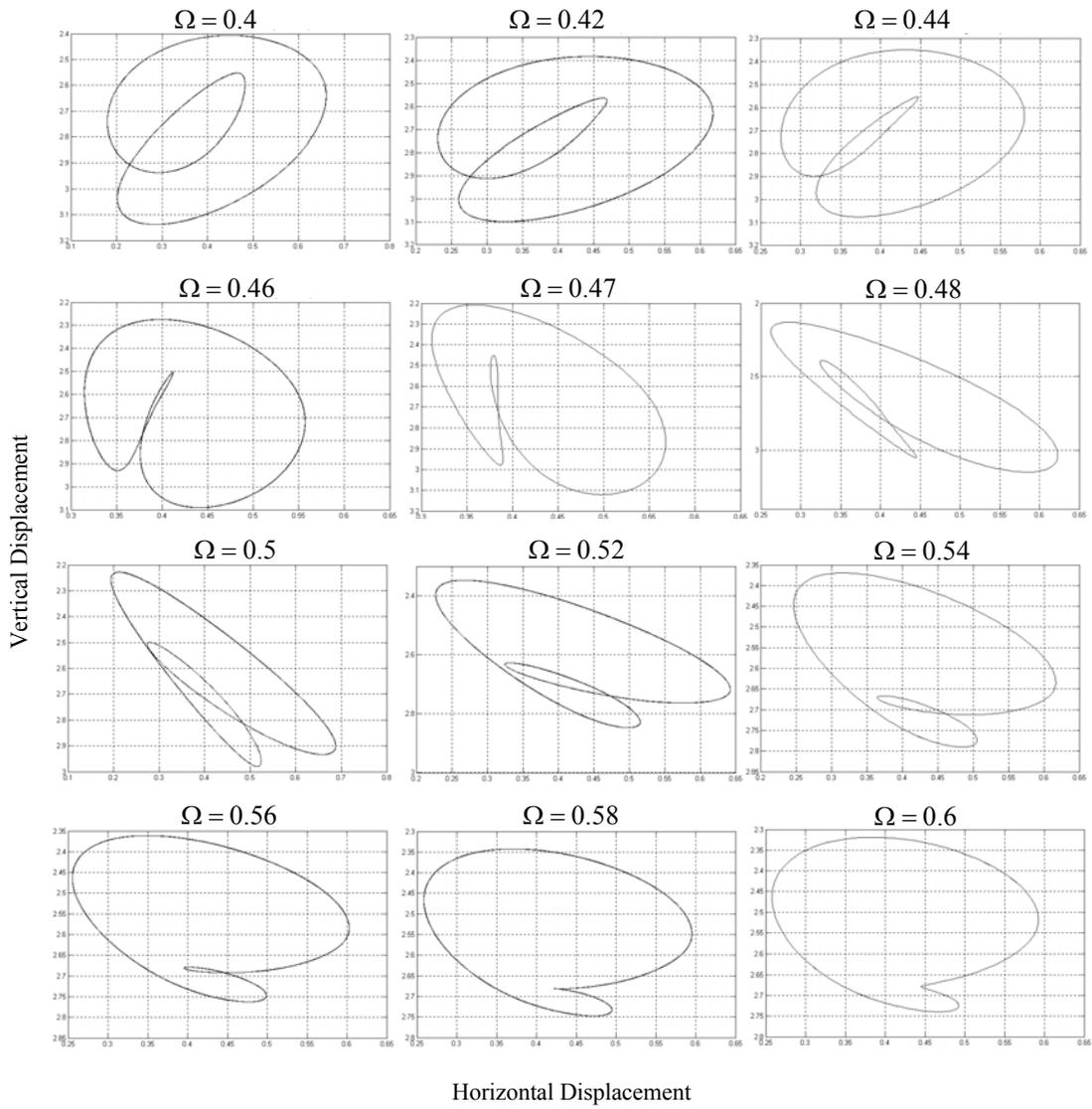

Figure 5: Orbital evolutions of the disc around half of the resonance
(Crack-imbalance orientation $\Phi_d$ =270°)

Figure 5: Evolutions des orbites au niveau du disque autour de la moitié de la résonance
(Orientation fissure-balourd $\Phi_d$ =270°)